%% file: main.tex
\documentclass[manuscript]{acmart}

\usepackage{subfig}
\usepackage{natbib}
\newcommand{\glmmci}[5]{$\beta$=#1, SE=#2, CI$_{95\%}$=[#3, #4], p#5}

\AtBeginDocument{%
  \providecommand\BibTeX{{%
    \normalfont B\kern-0.5em{\scshape i\kern-0.25em b}\kern-0.8em\TeX}}}

\copyrightyear{2021}
\acmYear{2021}
\setcopyright{acmlicensed}\acmConference[MuC '21]{Mensch und Computer 2021}{September 5--8, 2021}{Ingolstadt, Germany}
\acmBooktitle{Mensch und Computer 2021 (MuC '21), September 5--8, 2021, Ingolstadt, Germany}
\acmPrice{15.00}
\acmDOI{10.1145/3473856.3473863}
\acmISBN{978-1-4503-8645-6/21/09}

\begin{document}

\title[Comparing Concepts for Embedding Vocabulary Acquisition into Smartphone Interactions]{Comparing Concepts for Embedding Second-Language Vocabulary Acquisition into Everyday Smartphone Interactions}

\author{Christina Schneegass}
\email{christina.schneegass@ifi.lmu.de}
\affiliation{%
  \institution{LMU Munich}
  \city{Munich}
  \country{Germany}
  \postcode{80337}
}

\author{Sophia Sigethy}
\email{sophia@sigethy.de}
\affiliation{%
  \institution{LMU Munich}
  \city{Munich}
  \country{Germany}
  \postcode{80337}
}

\author{Malin Eiband}
\email{malin.eiband@ifi.lmu.de}
\affiliation{%
  \institution{LMU Munich}
  \city{Munich}
  \country{Germany}
  \postcode{80337}
}

\author{Daniel Buschek}
\email{daniel.buschek@uni-bayreuth.de}
\affiliation{%
  \institution{Research Group HCI + AI, Department of Computer Science, University of Bayreuth}
  \streetaddress{Universit\"atsstra{\ss}e 30}
  \city{Bayreuth}
  \country{Germany}
  \postcode{95447}
}

\renewcommand{\shortauthors}{C. Schneegass et al.}

\newcommand{\appstd}{StandardApp}
\newcommand{\appunlock}{UnlockApp}
\newcommand{\appnotif}{NotificationApp}

\begin{abstract}

We present a three-week within-subject field study comparing three mobile language learning (MLL) applications with varying levels of integration  into everyday smartphone interactions: We designed a novel (1) \textit{\appunlock} that presents a vocabulary task with each authentication event, nudging users towards short frequent learning session. We compare it with a (2) \textit{\appnotif} that displays vocabulary tasks in a push notification in the status bar, which is always visible but learning needs to be user-initiated, and a (3) \textit{\appstd} that requires users to start in-app learning actively. Our study is the first to directly compare these embedding concepts for MLL, showing that integrating vocabulary learning into everyday smartphone interactions via \textit{\appunlock} and \textit{\appnotif} increases the number of answers. However, users show individual subjective preferences. Based on our results, we discuss the trade-off between higher content exposure and disturbance, and the related challenges and opportunities of embedding learning seamlessly into everyday mobile interactions.    
\end{abstract}




\maketitle

\input{content/01_intro}
\input{content/02_related_work}
\input{content/03_Implementation}

\input{content/04_userstudy}
\input{content/05_discussion}
\input{content/06_conclusion}

\begin{acks}
This project is funded by the Bavarian State Ministry of Science and the Arts and coordinated by the Bavarian Research Institute for Digital Transformation (bidt).
\end{acks}
\bibliographystyle{ACM-Reference-Format}
\bibliography{bibliography}

\end{document}

%% file: content/01_intro.tex
\section{Introduction}
\label{sec:intro}
Language learning is a long-term task that requires users to engage with the learning content on a regular basis. Repeating content frequently can consolidate the memory of it and thereby counteract forgetting~\cite{schacter1999seven}. Prior research has explored a variety of options for technology to support this process. The concept of micro-learning utilises the presentation of small chunks of information, i.e. micro-content units, through micro-interactions and has proven its potential for effective vocabulary acquisition and recall~\cite{bruck2012mobile}. 
The sheer ubiquity of mobile devices makes them already a feasible tool for the presentation of learning content. We interact with our smartphones on a regular basis, frequently over the course of the day~\cite{ferreira2014contextual, hintze2017large}. 

Besides commonly available learning applications, research has evaluated concepts that give learning tasks more visibility during everyday smartphone interactions, such as presenting content as interactive push notifications~\cite{dingler2017language} or as lockscreen wallpapers~\cite{dearman2012evaluating}. These concepts aim to increase users' exposure to the learning content by maximising the visibility of the tasks~\cite{dearman2012evaluating} or by exploiting idle moments such as waiting situations~\cite{dingler2017language, cai2014wait}. While all these concepts are successfully evaluated individually, there has not yet been a comparative evaluation investigating different levels of embedding learning into everyday smartphone interactions. 

To address this gap, we conducted a three-week within-subject field study, which compares users' interactions and exposure to learning content using three learning apps that represent different embedding levels, described as follows: 

\begin{enumerate}
    \item First, our novel concept called \textit{\appunlock} extends the concept of Dearman and Truong's~\cite{dearman2012evaluating} learning wallpaper by connecting a simple multiple-choice vocabulary task that the user can answer by a single button press with the phone unlock action. Although users can still dismiss the learning task, this concept is strongly embedded into the unlock action as the app initiates the learning task. We aim to nudge users toward more frequent learning by lowering the threshold for users to interact with the content.

    \item Second, our notification learning app (\textit{\appnotif}) presents a learning task in a continuously shown notification. The notification increases learning task's visibility during interactions with the lockscreen or status bar but still requires users to actively initiate learning. 

    \item Third, we implement a \textit{\appstd}, following a baseline self-contained and self-initiated learning design. 
\end{enumerate}

In our in-the-wild evaluation, we investigate how the different degrees of embedding and self-initiated vs app-initiated exposures impact interactions with the learning content over the day and how users perceive this form of learning in their everyday use. 

The results show that the \textit{\appunlock} and \textit{\appnotif} designs, compared to the \textit{\appstd,} significantly increase the number of learning tasks users answer per day. The \textit{\appunlock}, in particular, further distributes the learning more evenly across the day. Interestingly, we observe a great variety in participants' subjective perceptions of the three concepts. While some people like that the \textit{\appunlock} nudges them to engage with the vocabulary regularly, others favored the \textit{\appnotif}, arguing that it is less distracting, especially when unlocking their phone with a specific goal in mind. In contrast, when participants had time to spare, they enjoyed long learning streaks with the \textit{\appstd}. 

Based on these comparisons, we discuss challenges and opportunities of embedding learning tasks more ubiquitously into peoples' everyday lives. In particular, combinations of features from the different concepts and offering further personalisation options can target individual user preferences and balance the trade-off between nudging users to learn and disturbing routine smartphone tasks.

%% file: content/02_related_work.tex
\section{Related Work}
\subsection{Mobile \& Micro-Learning}
\label{sec:relwork}
The ubiquity of mobile devices fosters the shift from learning being a classroom-based activity to an activity that is embedded into our daily life. By applying the micro-learning paradigm, learning content is broken down into small bits of information and presented to learners using simple interaction concepts~\cite{bruck2012mobile}. Hereby, the learning task is spread out over time. Especially for vocabulary acquisition, a higher number of short interactions should be favored over one long learning streak~\cite{cull2000untangling}. By spacing out the learning content, prior work has demonstrated that learners could memorise vocabulary content better than immediate successive presentation~\cite{dempster1987effects}. The concept of ``spaced-repetition'' suggests that repeating learning content according to a repetitive schedule helps consolidate the memory of it~\cite{ausubel1965effect, tabibian2019enhancing}. This schedule targets the transient forgetting of information that naturally occurs after a certain amount of time of the information is not rehearsed or applied~\cite{schacter1999seven}. As MLL apps require self-directed learning, it needs to be taken into account that the success of autonomous learning and the adherence to rehearsal schedules depends on the consideration of various dimensions (i.e. context of learners, learner interest and motivation, language proficiency, etc.)~\cite{kukulska2012chapter,lai2018self}. 
Thus, the overall goal of micro-learning is to maximise the degree of exposure with the learning content due to increased presentation and, thus, frequent repetition of content. In particular for the teaching of languages, the application of the micro-learning approach has proven to improve vocabulary acquisition and recall~\cite{cai2014wait,edge2011micromandarin, trusty2011augmenting}. Prior work presented different ideas for realizing micro-learning in mobile learning. For example, \citet{dearman2012evaluating} implemented a mobile language learning application that can display learning tasks (i.e. vocabulary translations with three multiple-choice options) on the lockscreen. The authors show that users frequently interacted with the application and improved their knowledge of the language over the course of this study. 
\citet{dingler2017language} achieved similar results by implementing and evaluating a vocabulary application that presents tasks in push-notification. Their app ``QuickLearn'' detects opportune moments for the presentation of the learning content, in particular, when users are bored and use their smartphone without a specific purpose. \citet{dingler2017language} concluded that their application increased the number of ``quick'' learning sessions on the go, which was appreciated by their participants.

\subsection{Smartphone Usage}
Users' interactions with their mobile devices can range in their level of intensity. Prior work has shown that half of our interactions with our smartphones take less than 30 seconds, with only one in ten interactions exceeding four minutes~\cite{yan2012fast}. In particular, many interactions do not even reach the threshold of fifteen seconds, what \citet{ferreira2014contextual} defined as so-called ``micro-usage'' situations. For example, users briefly reply to a message, check their phone for new push notifications, or want to know the time. 
For many of those interactions, users do not unlock their device \cite{hintze2014mobile, hintze2017large} or only use one application after unlocking~\cite{mahfouz2016android}. 
However, unlocking the mobile device is still an action that happens frequently throughout the day. \citet{mahfouz2016android} report that users unlock their phone on average 46 times, while other studies report an average of between 25~\cite{hintze2014mobile} and 47 unlocks~\cite{harbach2014sa} per day per participant. As this interaction has no other purpose than the authentication, prior work suggests combining the authentication event with a micro-interaction task, in particular, when the slide-to-unlock authentication is used. \citet{truong2014slide} emphasised that authentication gestures serve no purpose other than to unlock the phone and outline their potential to serve other tasks such as data collection. Therefore, it has been proposed as a quick and easy method to enter journaling data~\cite{zhang2016examining}, perform nutrition tracking~\cite{jung2017harnessing}, or sleep tracking~\cite{choe2015sleeptight}.

\subsection{Summary \& Research Gap} 
In summary, micro-learning applications for vocabulary-based language learning have proven to increase vocabulary recall \cite{cai2014wait,edge2011micromandarin, trusty2011augmenting} and have shown great potential for pervasively embedding learning into everyday smartphone usage~\cite{dearman2012evaluating, dingler2017language}.
Yet, it remains unclear how best to embed the learning task into everyday device use. This motivates our idea and investigation of our new concept of integrating learning directly into the authentication process, by displaying a vocabulary task every time the phone is unlocked using any form of authentication mechanism. 
Furthermore, to the best of our knowledge, no comparative analysis has been performed yet to investigate potential differences between the proposed pervasive presentation of vocabulary and a ``standard'' in-app learning application that requires the user to actively initiate learning. For example, while \citet{dearman2012evaluating} showed a knowledge gain through their wallpaper app, they do not implement a ``standard'' app as a control condition.

%% file: content/03_Implementation.tex
\section{Implementation}
We implemented three concepts with varying levels of integration following existing research ideas presented in Section~\ref{sec:relwork}, as Android applications (for Android 6.0 and higher). The \textit{\appunlock} (see Figure~\ref{fig:unlock}) presents vocabulary immediately following the unlock action.
A word is presented in a foreign language (L2) and the user is asked to choose between two presented translations in their mother tongue (L1) by pressing a button on the screen. We chose to implement a multiple-choice approach with two answer options as it enables quick and simple interaction with the content.

We further implemented a push notification learning app, which presents vocabulary in a pervasively displayed notification in the status bar (named \textit{\appnotif}, see Figure~\ref{fig:noti_bar}, along the concept proposed by \citet{dingler2017language}) and on the lockscreen (see Figure~\ref{fig:noti_lockscreen}). 

Finally, we implemented a baseline vocabulary app, similar to the interactions with established apps, which the user has to actively start and quit (named \textit{\appstd}, see Figure~\ref{fig:standard}).

Beyond the different presentations, a key conceptual difference is that the \textit{\appunlock} nudges the user to interact with the content, while the \textit{\appnotif} and \textit{\appstd} require users to actively initiate the learning themselves. However, in contrast to the \textit{\appstd}, the \textit{\appnotif}'s constant visibility acts as a continuous reminder for users to engage with the learning content. 

\subsection{Learning Content}
All three apps include vocabulary allowing for learning Spanish, French, and Swedish with translations into the native language of our University's country. The word list comprised of 446 common nouns (originally from the British National Corpus\footnote{http://www.natcorp.ox.ac.uk}) and have been translated to the other languages using Google Translate while removing inaccurate translations and highly ambiguous words, similar to the procedure of \citet{cai2014wait} and \citet{dingler2017language}. 

In the background, all three applications are connected to a Firebase database, which stores both the learning content and interaction logging data. In the database, each user is assigned a unique id to ensure anonymity. All three applications are described in more detail in the following.

\subsection{\appunlock}
\begin{figure*}[tb]
    \centering
        \subfloat[][UnlockApp]{
        \includegraphics[width=0.23\textwidth]{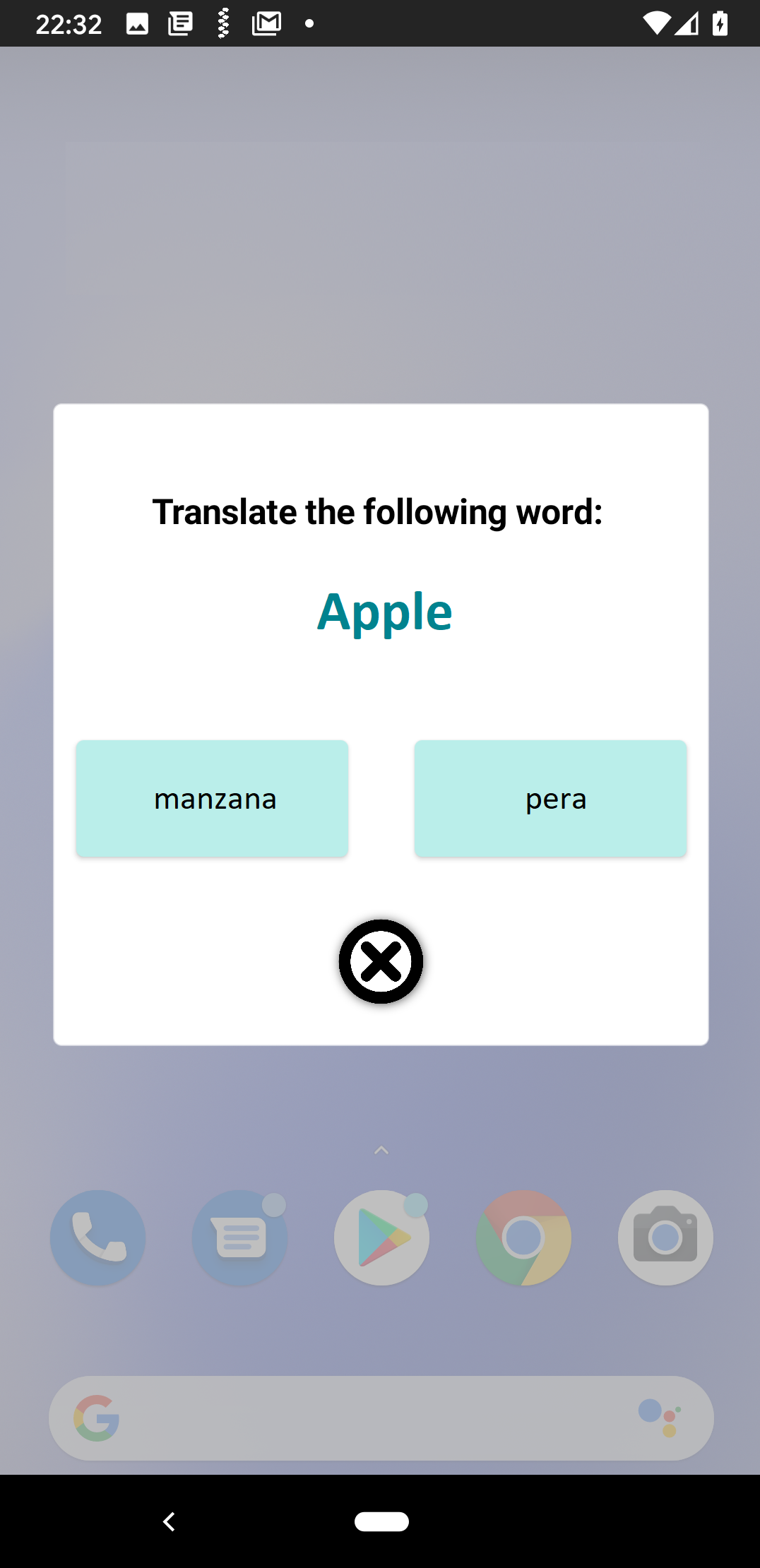}
        \label{fig:unlock}}
    \subfloat[][NotificationApp - \\Lockscreen View]{
        \includegraphics[width=0.23\textwidth]{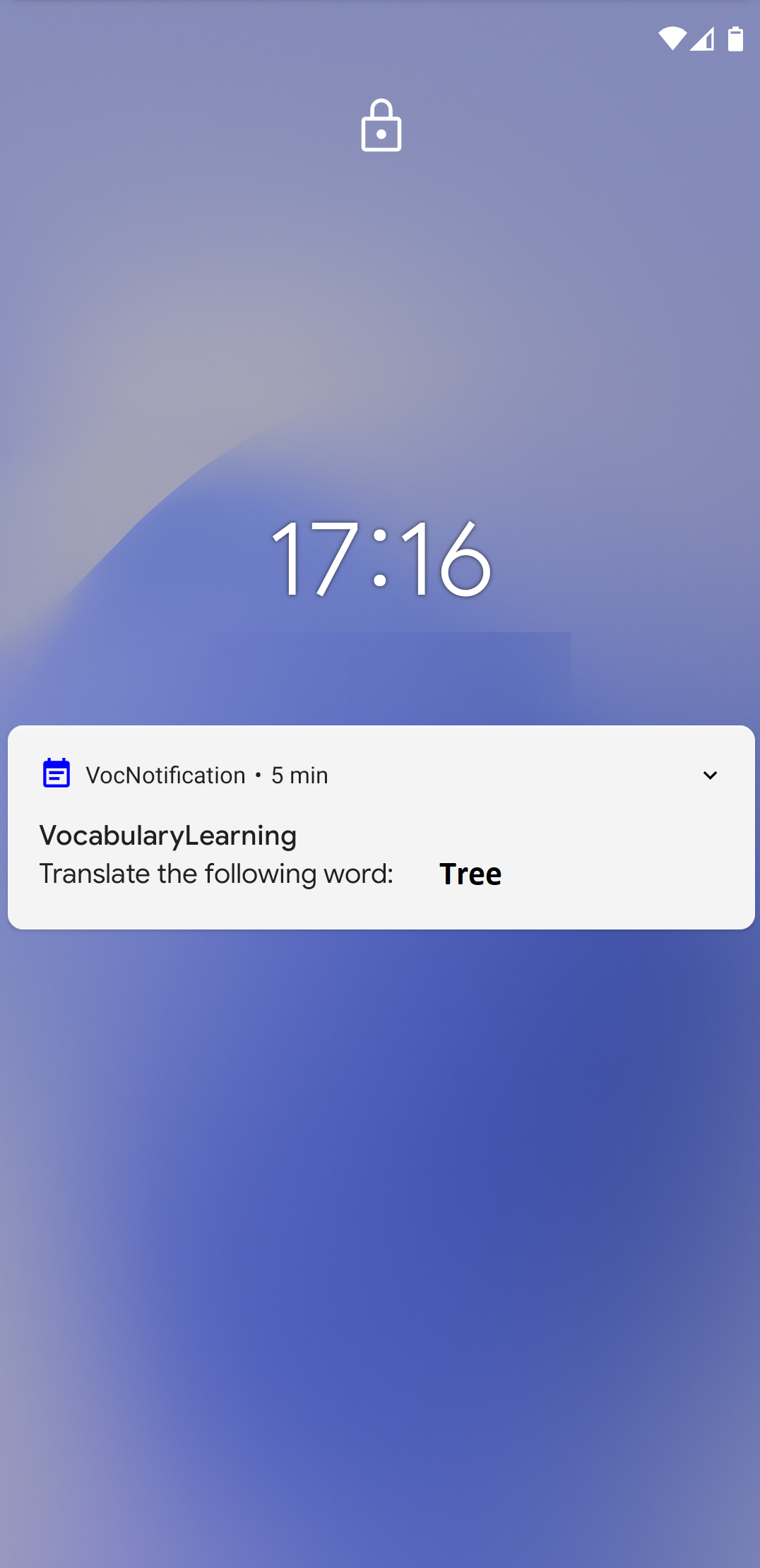}
        \label{fig:noti_lockscreen}}
    \subfloat[][NotificationApp - \\Status Bar View]{
        \includegraphics[width=0.23\textwidth]{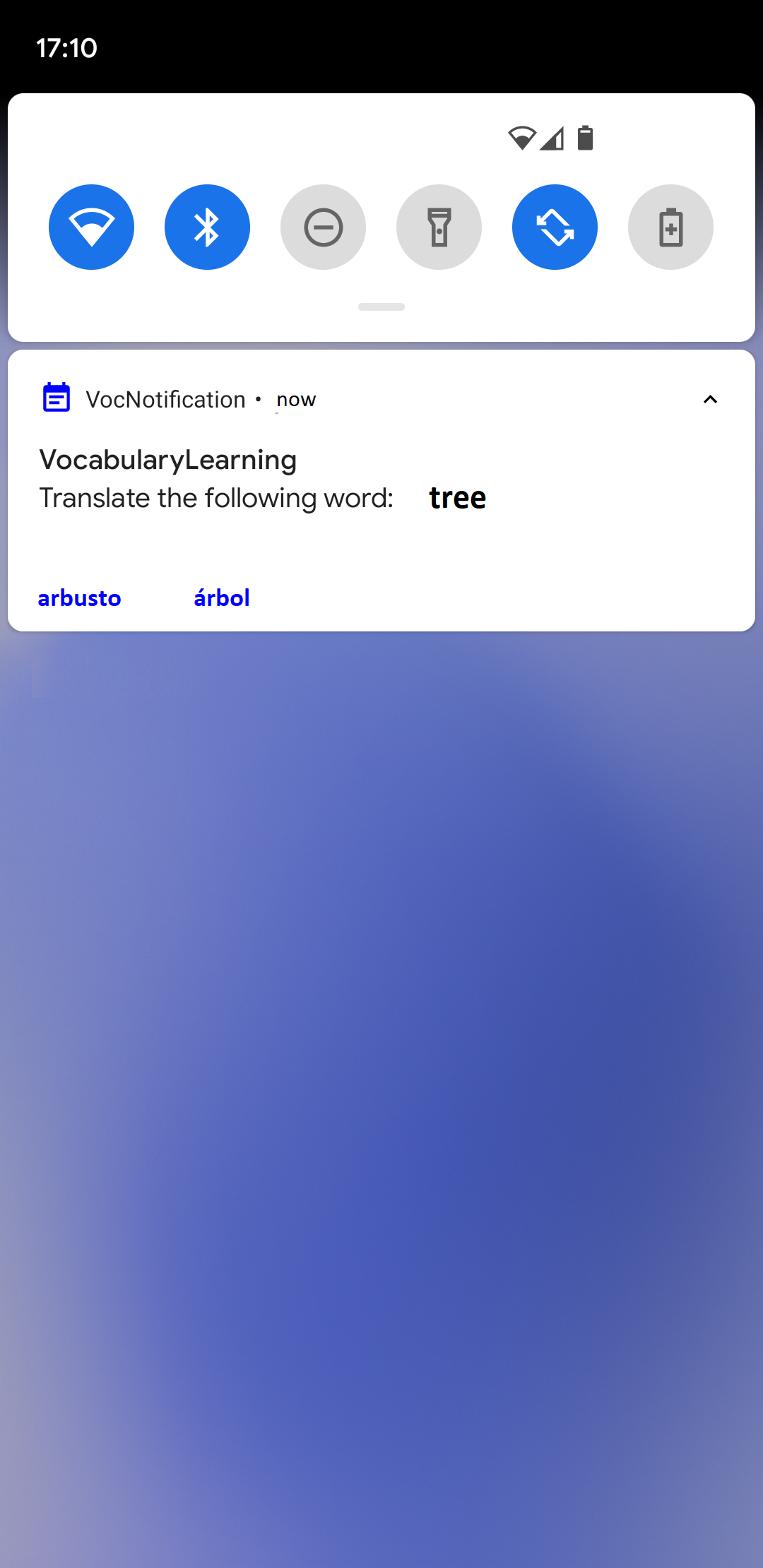}
        \label{fig:noti_bar}}
    \subfloat[][StandardApp]{
        \includegraphics[width=0.23\textwidth]{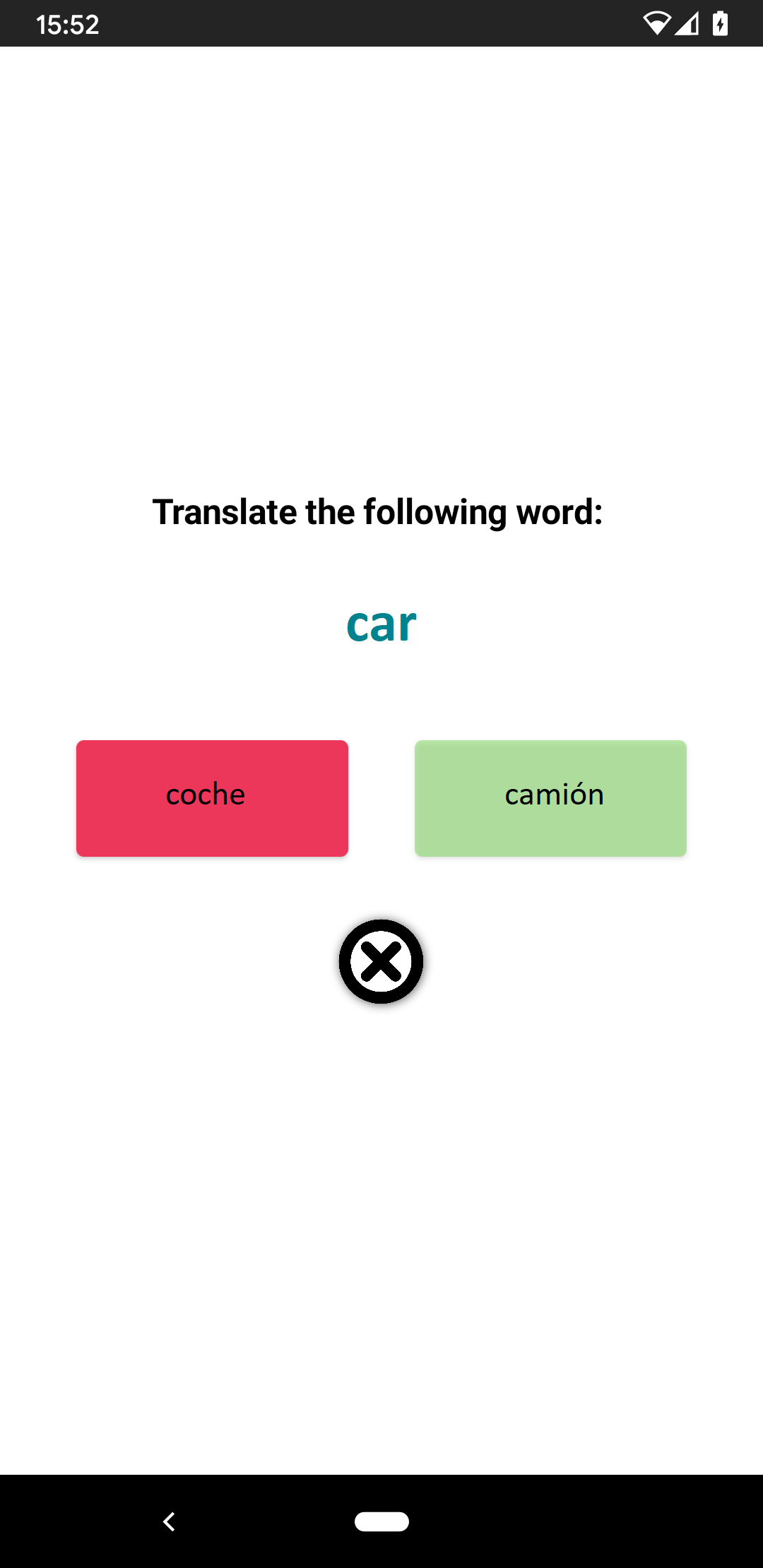}
        \label{fig:standard}}
    \caption{The \textit{\appunlock} (a) displays a vocabulary task immediately after the user performs the authentication. The \textit{\appnotif} (b+c) presents the learning task in a continuously displayed push notification. This notification is visible on the lockscreen (b) and in the (pulled down) status bar (c). Via notification action buttons (also displayed on the lockscreen if interactivity is enabled), the user can select one out of two translations for the respective word. Figure (d) shows the \textit{\appstd}, in the state of showing corrective feedback.}
    \label{fig:screenshots} 
\end{figure*}

To integrate the learning task into the authentication action, we display the vocabulary translation interface immediately when the user unlocks the screen (see Figure~\ref{fig:unlock}). As the current Android versions do not allow a direct manipulation of the lockscreen, we chose to implement a context-registered broadcast receiver. This receiver listens for the unlock event (ACTION\_USER\_PRESENT indicates that the user is present after the device is woken up\footnote{Android User Intent: \url{https://developer.android.com/reference/android/content/Intent}, last accessed February 4th, 2021}) while running continuously in the background and places the app in the foreground when an unlock occurs. Since it is possible that applications are shut down for battery optimisation purposes, we further included a foreground service that allows for automatic restart of the application. 

When an unlock event occurs, the \textit{\appunlock} asks the user to translate one word from the native language to one of the three languages of choice -- Spanish, French, or Swedish. To keep the learning task as short as possible, the interface only presents the translation task, two potential translation options in the form of buttons. Further, we include a ``skip'' option to keep learners from guessing the answer, as guessing increases the risk of experiencing the negative testing effect \cite{marsh2009memorial} (i.e. remembering the false answer one gave to a multiple-choice question rather than the correct solution). 

After the user selects one of the two translation options, the \textit{\appunlock} displays corrective feedback by highlighting the answer option in either red or green. Providing feedback in multiple-choice learning tasks is essential and counteracts the negative testing effect~\cite{butler2008feedback}. After a short delay for users to perceive the feedback, the app moves into the background.

We chose a translucent background to not distract the user too much from the primary task they were about to accomplish when unlocking the phone. Prior research by \citet{hodgetts_contextual_2006} has shown that when the primary task was still visible during an interruption, participants are able to resume the task faster. To ensure consistency, we used the Android material design and color palettes. 

The \textit{\appunlock} further allows to use a fingerprint sensor gesture (swiping left or right on the fingerprint sensor) as input to answer the tasks. We utilise the Fingerprint Gesture Controller accessibility service\footnote{FingerprintGestureController \url{https://developer.android.com/reference/android/accessibilityservice/FingerprintGestureController}, last accessed February 4th, 2021}. This controller registers gestures performed on the fingerprint sensor, independent from sensor location (depending on the model, the sensor can be located on the back of the phone, on the side, or on the front below or on the screen). While prior work has already shown that fingerprint sensor gestures can be a feasible tool for quick input such as unlock journaling \cite{fortin2019exploring}, this feature is not yet supported by many Android devices. Thus, we will not evaluate the usage of it in this study.

\subsection{\appnotif}
In this app, learning content is presented in a continuously displayed push notification, both in the (pulled down) status bar of the smartphone (see Figure~\ref{fig:noti_bar}) but also on the lockscreen (see Figure~\ref{fig:noti_lockscreen}). This enables the user to interact with the content at any point in time and answer as many tasks as they like to, the continuous notification working as an active and constant reminder. The notification displays the task and word to be translated as well as two potential translation options via action buttons\footnote{Android Notification Action Buttons: \url{https://developer.android.com/training/notify-user/build-notification\#Actions}, last accessed February 4th, 2021}. Those buttons are frequently applied by other applications to allow immediate interaction with the app sending the notification. For example, Gmail\footnote{Gmail: \url{https://www.google.com/gmail/about/}, last accessed February 4th, 2021} employs action buttons to enable users to delete or reply to emails directly from the notification. We implemented a broadcast receiver to listen for the input in the notification and check the correctness. In the same style as the \textit{\appunlock}, green and red coloring of the two potential translations indicates corrective feedback. Afterward, the notification is updated and a new word is shown. Likewise, when the user dismisses a notification or restarts the app, it presents a new word. The notification itself cannot be removed by the user but remains constantly visible in the pulled down status bar. 

\subsection{\appstd}
In contrast to the \textit{\appunlock} and \textit{\appnotif}, the \textit{\appstd} does not show any continuous reminder of the learning task. The users have to actively open the application and engage with the learning content. Once the users open the app, they can complete as many learning tasks (i.e. request as many translation tasks) as they like. The appearance of the application's interface is consistent with the interface of the \textit{\appunlock}: The app shows the native word, the two translation options in buttons, a dismiss button, and corrective feedback for the task (see Figure~\ref{fig:standard}). In contrast to the \textit{\appunlock}, the background is fully opaque.

%% file: content/04_userstudy.tex
\section{Evaluation}
Our evaluation follows a within-subject design, where all participants interact with all three different applications to allow them to draw comparisons. Each app is used over the course of one week (seven days) before switching to the next app. The vocabulary progress is preserved so that people can continue in the next app where they left off in the previous one. The order of the apps is counter-balanced across participants to avoid sequence effects.

In our subsequent analysis, we explore differences in users' interaction with the three apps (independent variable), in particular frequency and duration of interactions, as well as perceived usability and experience (dependent variables). Based on related work, we particularly evaluate differences among the three apps with regard to how much they expose users to the learning content, stating the following main hypothesis:
\begin{itemize}
    \item[$H_1$] Embedding vocabulary tasks into everyday smartphone interactions as in the \textit{\appunlock} and \textit{\appnotif} leads to a higher number of learning tasks solved by users over the course of the study compared to the \textit{\appstd}.
\end{itemize}

Furthermore, we gather individual feedback on participants' preferences and experiences when learning with the three applications.

\subsection{Procedure}
We provided participants with a detailed installation guide enabling people to download the three applications from the Google Play Store (setup as a non-public test version), after they gave informed consent. The guide also encouraged them to restart the (current) app in case they restart their phones or the app is closed for any other reason and does not restart automatically.

At the beginning of the study, people filled in a questionnaire on demographics and current authentication method, smartphone usage habits, language proficiency, and motivation to learn a new language. In this process, people selected a language to learn with the apps (Spanish, French, or Swedish). We informed participants that the vocabulary taught in the apps is on a beginner's level and thus recommended to choose a language they are not very proficient in.

During the study, we logged data on the users' interaction with the apps, including the number of solved vocabulary tasks, their correctness, response times and task dismissals (in the \textit{\appunlock}).

\subsection{Sample}
We recruited 30 participants via University mailing lists, social media channels, and word-of-mouth, nineteen identifying as female and eleven as male. Their average age was 29.8 years ($SD=15.55$, range 19-78 years) and the majority (18) reported to be studying or working full time in jobs such as Engineer, Physicist, Accountant, Doctor, or Dentist, while two were pensioners.
Fifteen stated having a high school degree and twelve a university degree (including bachelor, master, or phd). One third (19 people) reported using fingerprint authentication on their smartphones, and seven reported using a PIN or password, two patterns, and two face recognition. All participants stated to use their device at least multiple times a day (15) if not multiple times per hour (15). 

At the time of the study, ten participants confirmed learning or actively improving on a language (four Spanish, three French, two English, and one Japanese). They further stated to be proficient in at least one and up to five foreign languages ($Md=3$), and reported to have great interest in learning a new language ($M=6.07$, $SD=0.85$, Likert-scale from 1=``I fully disagree'' to 7=``I fully agree''). 
As languages to learn in this study, thirteen people chose Spanish, twelve French, and five Swedish. For their successful participation all people received a 25 Euro Voucher for an online store, or an equivalent amount of study credit points.

\subsection{Results}
We report on the data of 30 people. However, one person did not complete any of the final questionnaires, and two additional people did not submit the questionnaire for the \textit{\appstd}. Therefore, we report the interaction data of 30 and the questionnaire results of 29 people for the \textit{\appunlock} and \textit{\appnotif}, 27 for the \textit{\appstd}, respectively. For the statistical analysis of the questionnaires, we performed pair-wise exclusion for incomplete data sets.

We used R~\cite{R2020} for significance testing; concretely, (generalised) linear mixed-effects models (LMMs, packages \textit{lme4}~\cite{Bates2015} and \textit{lmerTest}~\cite{Kuznetsova2017}). The LMMs accounted for individual differences (\textit{participants}) and \textit{app order} via random intercepts. Note that app order was counterbalanced yet we still included it here following best practices. As fixed effects, we included \textit{app}, plus the \textit{number of days} since the start of the study. 

\begin{figure}[tb]
    \centering
        \includegraphics[width=1\textwidth]{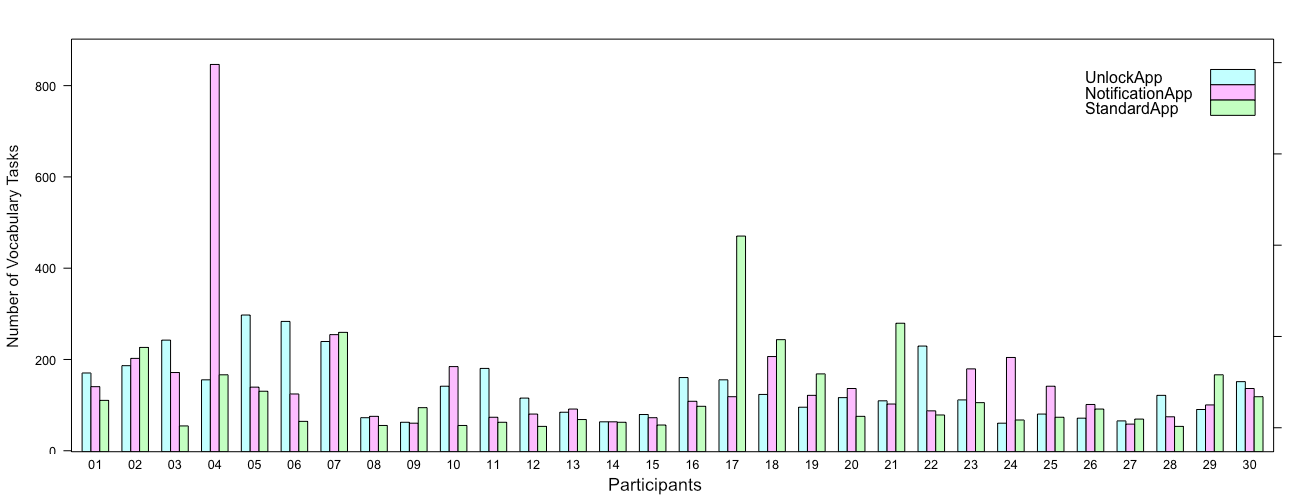}
        \caption{Overview of participants' usage of the three apps during the study: (per participant from left to right) the overall number of vocabulary tasks solved in the \textit{\appunlock}, \textit{\appnotif}, and \textit{\appstd}.}
    \label{fig:histogram} 
\end{figure}

\subsubsection{Overall Usage}
In total, we recorded 7715 answered vocabulary tasks over all participants across the three weeks of our study. The most tasks were answered using the \textit{\appnotif} (2945), followed by the \textit{\appunlock} (2604) and the \textit{\appstd} (2166). On average per day, people used the apps to solve 10-13 vocabulary tasks. Here, the \textit{\appstd} showed the lowest mean usage ($Md=2$, $M=10.28$, $SD=20.05$), while the \textit{\appunlock} ($Md=8$, $M=12.28$, $SD=14.73$) and \textit{\appnotif} ($Md=5.5$, $M=13.7$, $SD=28.43$) were used more frequently. While the \textit{\appnotif} shows the highest overall usage, the Median is lower than the Median of the \textit{\appunlock}. This is due to one exceptional case in the \textit{\appnotif} usage (see P4 in Figure~\ref{fig:histogram}): P4 reported having used the \textit{\appnotif} extensively out of enjoyment, with over 800 answered vocabulary tasks (increasing the overall usage count but with little effect on the Median). Since this was intended use, we do not remove this as an outlier from our analysis.

For significance testing, we fitted a generalised LMM (Poisson family) on the answer count data (i.e. number of answered vocabulary questions)\footnote{We will report the results of the GLMM in a concise format (cf.~\cite{Meteyard2020}) and provide the full analysis in the supplementary material.}.
The model had \textit{app} as a significant positive predictor (\textit{\appunlock}: \glmmci{.43}{.06}{.31}{.55}{<.0001}; \textit{\appnotif}: \glmmci{.36}{.06}{.25}{.48}{<.0001}): Therefore, compared to the \textit{\appstd}, using the \textit{\appunlock} was estimated by the model to result in $\exp(\beta)=1.54$, that is, $54\%$ more answered vocabulary questions. Similarly, using the \textit{\appnotif} was estimated to result in $43\%$ more.
Moreover, the model had \textit{day} (since start of the study) as a significant negative predictor (\glmmci{-.15}{.01}{-.17}{-.12}{<.0001}): The number of answered questions was estimated by the model to decline over the course of the study (estimated as $\exp(\beta)=0.86$ i.e. $-14\%$ per day).
The interaction of \textit{day} and \textit{app} was significant for \textit{\appunlock} (\glmmci{-.07}{.02}{-.10}{-.04}{<.0001}), but not for \textit{\appnotif} ($p=.176$).

\begin{figure}[tb]
    \centering
        \includegraphics[width=1\textwidth]{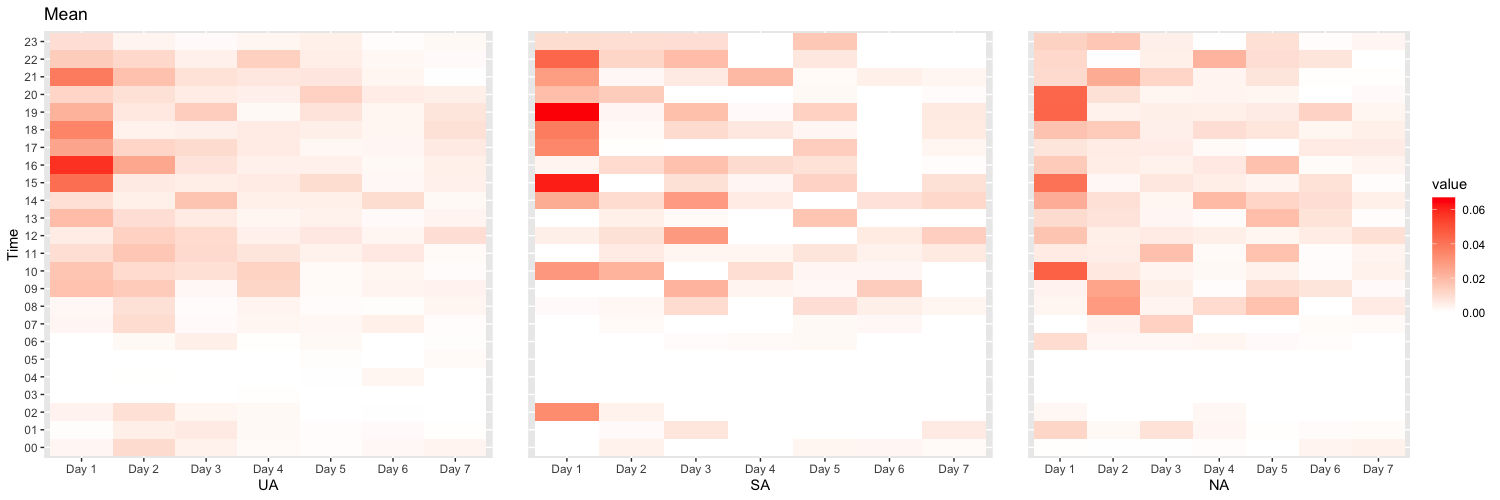}
        \caption{The relative distribution of vocabulary tasks solved for the \textit{\appunlock} (left), \textit{\appstd} (middle), and \textit{\appnotif} (right) in percent (0.06 = 6\%). The amount of vocabulary tasks is normalised for each user with respect to their overall number of solved tasks and visualised according to the seven usage days (x-axis) and the hours of a day (midnight to midnight, y-axis).}
    \label{fig:heatmap_tod} 
\end{figure}

We further plotted people's interactions with the three applications over the course of the whole day and the week of use: Figure~\ref{fig:heatmap_tod} visualises the number of tasks solved by the users, indicating that they interacted with each app most frequently on the first day of use. It has to be noted here that participants started the study on different days of the week (Monday: 5, Tuesday: 2, Wednesday: 2, Thursday: 5, Friday: 5, Saturday: 6, Sunday: 5) and the starting weekday then stayed the same for each person across app conditions.
Moreover, the plots reveal that interactions with the \textit{\appunlock} are scattered more across the time of the day compared to the interactions with the \textit{\appstd} and \textit{\appnotif}. Furthermore, the plot matches the results of the LMM that the number of answered questions declined from day one to day seven.

\subsubsection{Learning Task Correctness}
In general, the users answered more than 90\% of the learning tasks correctly over all three applications (see Table~\ref{tab:answer_Correctness}). The correctness rates for the \textit{\appstd} and \textit{\appnotif} are slightly higher than the rates for the \textit{\appunlock}. A Friedman test revealed no significant difference ($p>.05$) for the task correctness of the answers with regard to the three applications. 

For the \textit{\appunlock}, we can further measure the time between when the learning task is presented and the moment the answer is recorded. On average, it took participants 3.4 seconds to answer a task using the \textit{\appunlock} ($SD=1.02$). The time to complete the task for incorrect answers is higher ($M=3.98$, $SD=2.71$) compared to correctly answered tasks ($M=3.38$, $SD=1.02$). 

The \textit{\appunlock} enabled users to dismiss a learning task in case they did not want to answer it. During the seven study days, users chose to skip between zero and 80 tasks, with an average of 9.6 skips ($SD=14.48$, $Md=4$).

\begin{table}
    \centering
    \caption{Overview of people's use of the three apps in terms of the number of answered vocabulary questions.}
    \begin{tabular}{lcccc} 
    \toprule
    \textbf{App Type} & \textbf{\# Answers Total} & \textbf{\# Correct} & \textbf{\# Incorrect} & \textbf{\% Correct}\\
        \midrule
       \appstd & 2166 &  2026 & 140 & 93.54\% \\
        \appnotif & 2945 & 2762 & 183 & 93.79\% \\
        \appunlock & 2604 & 2380 & 224 & 91.40\% \\
        \bottomrule
    \end{tabular}
    \label{tab:answer_Correctness}
\end{table}

\subsubsection{Favorites, Ratings \& Subjective Results}
We asked participants directly which of our three applications they would prefer for everyday use. As their most favorite application, 18 out of 30 people named the \textit{\appnotif}, followed by seven mentions of the \textit{\appunlock}. Vice versa, 15 participants stated the \textit{\appunlock} as their least favorite of the three applications, followed by 13 naming the \textit{\appstd}.

\begin{figure*}[tb]
    \centering
        \subfloat[][Like]{
        \includegraphics[width=0.24\textwidth]{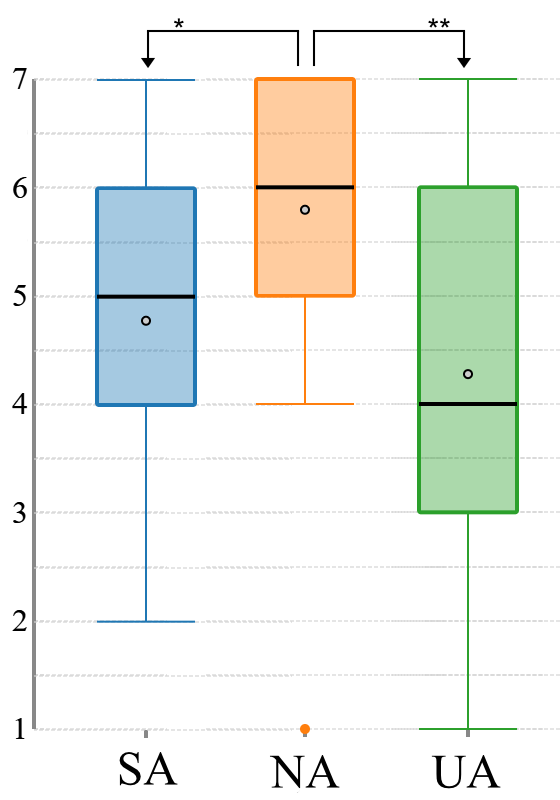}
        \label{fig:plot_like}}
    \subfloat[][Used When Time]{
        \includegraphics[width=0.24\textwidth]{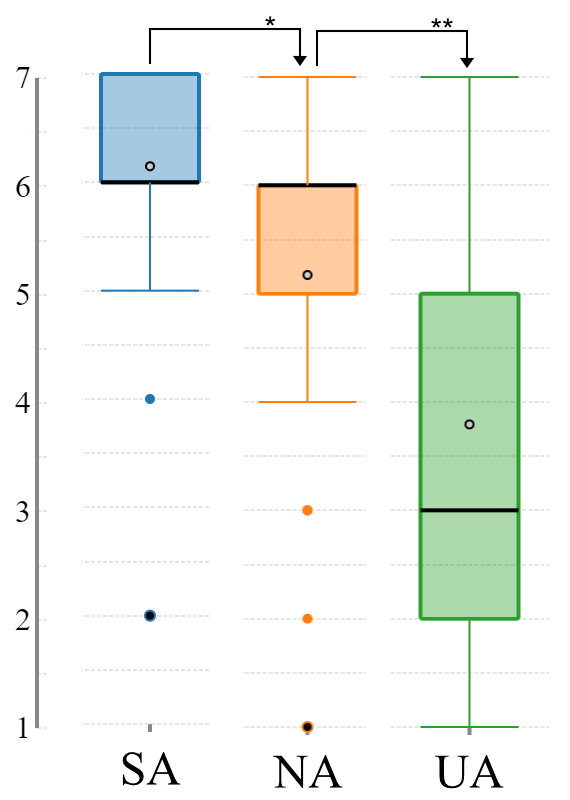}
        \label{fig:plot_when_time}}
    \subfloat[][Used Even Without Time]{
        \includegraphics[width=0.24\textwidth]{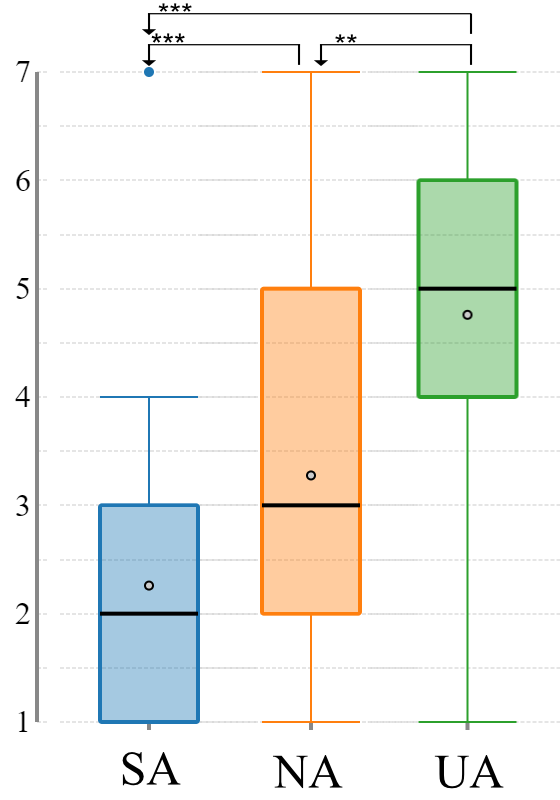}
        \label{fig:plot_even_without_time}}
    \subfloat[][Continue for Everyday Use]{
        \includegraphics[width=0.24\textwidth]{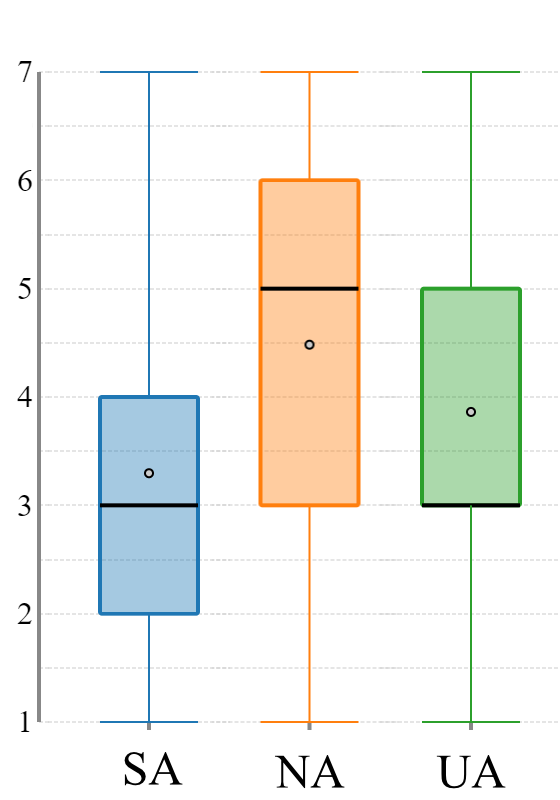}
        \label{fig:plot_everyday_usage}}
    \caption{Partcipants' ratings of the \textit{\appstd} (SA), \textit{\appnotif} (NA), and \textit{\appunlock} (UA) for four questionnaire items (from 1= ``I totally disagree'' to 7= ``I totally agree''): (a) I liked the [SA/NA/UA]; (b) I only answered tasks with the [SA/NA/UA] when I had the time to learn; (c) I answered tasks with the [SA/NA/UA] even when I did not have time to learn; and (d) I would continue using the [SA/NA/UA] frequently in my everyday life for learning a language. Asterisks indicate all statistically significant differences (* <.05| ** <.01| *** <.001).}
    \label{fig:plots} 
\end{figure*}

Moreover, we analysed participants' subjective impressions of the three applications (expressed in the final questionnaire as 7-point Likert scale items from 1= ``I totally disagree'' to 7= ``I totally agree''). For each item we performed a Friedman test with post-hoc Wilcoxon pairwise-comparisons. 
There was a significant difference in how much participants liked the applications (a: $\chi^2(2)=6.812$, $p<.05$, see Figure~\ref{fig:plot_like}). Post-hoc analysis with Wilcoxon signed-rank tests showed a significantly higher score for the \textit{\appnotif} compared to both the \textit{\appstd} ($Z=-2.159$, $p<.05$) and the \textit{\appunlock} ($Z=-3.100$, $p<.01$).

We further asked participants (b) if they felt they used the app only when they had time to learn and (c) if they used the app even when they had no time to learn. Both items revealed significant differences in participants' subjective perception of their usage among all three applications (b: $\chi^2(2)=16.568$, $p<.001$, see Figure~\ref{fig:plot_when_time}; c: $\chi^2(2)=15.571$, $p<.001$, see Figure~\ref{fig:plot_even_without_time}). For item (b), participants report to be more likely to use the \textit{\appstd} when they had time to learn compared to the \textit{\appnotif} ($Z=-2.060$, $p<.05$) and the \textit{\appunlock} ($Z=-3.960$, $p<.001$), and more likely to use the \textit{\appnotif} when compared to the \textit{\appunlock} ($Z=-2.737$, $p<.01$).
For item (c), participants report to be more likely to use the \textit{\appunlock} even when they had no time to learn compared to the \textit{\appnotif} ($Z=-2.786$, $p<.01$) and the \textit{\appstd} ($Z=-3.945$, $p<.001$), and more likely to use the \textit{\appnotif} when compared to the \textit{\appstd} ($Z=-3.945$, $p<.001$).

We found no significant differences in participants' ratings of the design of the three applications and their intuitiveness of use. Moreover, the Friedman test revealed no significant difference in terms of participants' willingness to continue using the applications in their everyday life. The descriptives, however, show a slight preference in favor of the \textit{\appnotif} ($M=4.37$, $SD=1.85$) compared to the \textit{\appunlock} ($M=3.70$, $SD=2.05$) and the \textit{\appstd} ($M=3.30$, $SD=1.73$), see Figure~\ref{fig:plot_everyday_usage}.

\subsubsection{Open Comments and Suggestions for Improvement}
Regarding the concept of the \textit{\appnotif}, people reported positive as well as negative impressions: One participant considered the continuous presentation not optimal and rather wants to take the time to actively engage with language learning (P15). P7 perceived the display of the notification on the lockscreen as distracting. Two other participants suggested increasing the delay between the presentation of two vocabulary tasks so that the app is not running all the time but just presents a new word from time to time (P21, P27). On the other hand, some participants positively emphasised the unobtrusiveness of the notification app, saying it is ``[...] not distracting, it provides the opportunity to quickly solve a couple of learning tasks'' (P11) and that it is ``very suitable for daily use'' (P25). To further improve the application, participants wished for the inclusion of grammar knowledge (P2, P5, P14), free text entry of words to foster recall rather than recognition of words (P2), audio output to help with the pronunciation (P5, P8, P14), and gamification features (P6).

Similar suggestions were stated for the \textit{\appunlock}: Participants wished for free text entry opportunities (P22), pronunciation support (P9, P26), and grammar knowledge (P3, P5, P7, P19). Since vocabulary in itself is not sufficient to learn a new language, P6 considers the app a nice addition for people currently attending language learning courses and P19 recommends using it to freshen up a language. 
Regarding the overall concept of combining the learning task with authentication, many participants stated positive impressions: While P9 highlighted the simple design, P19 emphasises that ``it's good that you always have to solve at least one task and therefore learn continuously''. Further, P6 stated to like the app and in particular the idea, similar to P8. However, P8 adds that ``[...] it bothers me that I have to answer the task or dismiss the app when I just quickly want to do something on my phone''. P20 shares the experience and describes that when they just quickly want to use the phone, they are ``not focused enough to answer the questions conscientiously''.
To address this issue, other participants suggested including more personalisation options. For example, P7 wishes for a feature to adjust the number of vocabulary tasks presented at each authentication  (which was fixed to one in the study), describing that if a user currently has more time to learn, then they could increase the number of words presented per unlock event. Further, P15 proposes to define time intervals during the day in which the app is active (P15).

%% file: content/05_discussion.tex
\section{Discussion}
\label{sec:discussion}

\subsection{Limitations}
Our user study was conducted during the COVID-19 pandemic, probably leading to anomalies in users' daily routines, mobility, and smartphone usage due to lockdown and work-from-home phases. All these factors potentially influence the use of mobile language learning applications; thus, we can not ensure the generalisability of our results. However, we included a control condition via the \textit{\appstd}. 

The focus of our evaluation lies on the users' experiences and users' interaction with the three apps of varying integration into smartphone usage. Although our applications do not constitute an exhaustive representation of all MLL applications, we chose these three as a concise comparison of how different levels of embedding could be implemented. Further, with our focus on user's interaction we can not report on actual vocabulary retention. However, as retention of vocabulary is increased by frequent interaction with the content \cite{cull2000untangling, dempster1987effects}, we expect our applications to positively impact people's vocabulary recall and recognition. Nonetheless, actual vocabulary retention, particularly concerning the consolidation properties of the apps over time, needs to be further evaluated in future work. 

By providing people the three applications to use for each seven days during this user study, we gained interesting insights into users' learning behavior. We decided in favor of a shorter within-subject study over a longer between-subject study to enable users to draw comparisons among their interaction with the three apps and to make the best use of the limited sample size. However, we are aware that our study only presents a narrow view onto users' actual behavior. As the users show a great diversity in individual preferences, only a long-term evaluation with a larger sample will be able to show which behaviours will prevail in users' daily lives. Our study data shows a more extensive usage of each application on the first day, declining over the seven days of usage. This pattern could indicate a form of curiosity or novelty effect. Educational technology research has shown that accustomisation with new learning technology can negatively impact learners' preferences regarding technology-based learning~\cite{krendl1992student} and reduce users' motivation~\cite{keller2004learner}. In contrast to the \textit{\appstd}, the \textit{\appunlock}'s concept partly counteracts this decline by continuously engaging the learner in solving vocabulary tasks with each authentication event. However, a future long-term evaluation in particular of the new \textit{\appunlock} concept is required to reveal the strength of the novelty effect in everyday usage.

\subsection{UnlockApp and NotificationApp Increase Vocabulary Exposure}
Our analysis confirms our initial hypothesis, showing that the \textit{\appnotif} and \textit{\appunlock} result in a higher number of solved vocabulary tasks per participant. Thus, we conclude that the learning content exposure was higher for these two apps when compared to the baseline, the \textit{\appstd}. Furthermore, the presentation of the vocabulary tasks after the authentication leads to a more spread out exposure to the tasks across the day, confirming \citet{dingler2017language}'s conclusion that proactively triggering vocabulary learning sessions can effectively space out learning. Further, the continuous presentation of content through the \textit{\appnotif} allows for short but also extensive learning sessions (cf. the outlier P4). The similar distribution of correctly and incorrectly answered tasks across the three applications gives no indication that users might have been less focused when learning with the \textit{\appunlock} or \textit{\appnotif}. Additionally, we observed that users occasionally skip tasks with the \textit{\appunlock}. This suggests that even though the skipping requires the same amount of effort as selecting an answer (one button press), the users can judge if they have the mental capacity and/or time to engage in the learning or not.

\subsection{Potential for Adaptation \& Personalisation}
The subjective feedback revealed individual differences regarding people's attitudes toward the three applications: While many state to like the concepts of the \textit{\appnotif}, opinions on the \textit{\appunlock} are mixed. Specifically, some people felt distracted by the \textit{\appunlock}'s vocabulary presentation when they unlocked their phone with a specific task in mind. People do not oppose the concept in general but express their need for further personalisation. Suggested adaptation features include the definition of learning time frames for the \textit{\appunlock} or adjusting the number of tasks presented with each authentication event. Moreover, we see potential for automated mechanisms that learn from people's interactions with the \textit{\appunlock} (in particular dismissals) and adjust the presentation accordingly. I.e. the app should stop promoting users with learning tasks after unlock events at times during the day when they are frequently dismissed.

\subsection{Extending the \appunlock{} Concept to Different Authentication Methods}
Based on the results here, we deem it useful to discuss possible extensions of the \textit{\appunlock} concept to further authentication methods: Concretely, the majority of participants in our study used fingerprint authentication to unlock (yet with phones that do not support fingerprint gestures). Our app and concept already support fingerprint gestures to respond to vocabulary tasks (see implementation section), which can help to embed the learning task even more implicitly into the authentication (i.e. no switch from fingerprint to touch required). Prior work has found that using fingerprint sensor gestures for journaling is perceived faster and less intrusive compared to notifications \cite{fortin2019exploring}. Both touch input and fingerprint gestures might also be combined with unlocking methods that do not require further input themselves, such as face unlock, or when Android's smart screen lock feature is enabled (i.e. no unlock at home). 
Beyond this, we already gathered experiences with knowledge-based authentication methods: One third of our sample stated to use PIN, password, or pattern authentication. In these cases, the \textit{\appunlock} also presents the task immediately after the unlock event.

\subsection{Extending the Learning Features of Embedded Learning Applications}
Our application focused on the presentation of vocabulary translations to reduce the complexity of the content. With this, we aim to simultaneously increase the control of potential effects among the three apps. The participants of our study strongly emphasised the demand for additional features such as grammar knowledge or pronunciation exercises in open comments, and compared our application to applications available on the market. To accommodate for people's need for more complex learning tasks, we see three potential solutions: (1) By combining the embedding concepts with a fully-featured learning app, we could address users' preferences for long learning streaks when they have time to spare (i.e., with the \textit{\appstd}), during which the app could also teach grammar or pronunciation knowledge. The differentiation between two complexity levels could also be realized by (2) differentiating between short and long learning sessions as suggested by \citet{cai2014wait} (e.g., by exploiting different types of waiting situations).  Additionally, the \textit{\appunlock} could refresh users' vocabulary knowledge by continuously presenting translation tasks after each authentication to engage users with a language on a daily basis.
Option (3) is to investigate extensions of learning tasks that are short enough to be embedded into the \textit{\appunlock} or \textit{\appnotif}. These tasks need to be solvable with simple interactions and with low effort and time. Possible examples include fill-in-the-blank tasks offering a word in two tenses or with two suffix options.

%% file: content/06_conclusion.tex
\section{Conclusion \& Future Work}
\label{sec:conclusion}
We have compared three basic concepts for embedding vocabulary learning tasks into everyday smartphone use. When similar ideas were proposed in the literature, these individual concepts had been evaluated independently of each other. Thus, in this paper, our main contribution to the literature is their direct empirical comparison in a single study.

In this light, our key result and conclusion is that embedding vocabulary learning tasks into everyday smartphone interactions -- concretely: the status-bar notifications or the unlock procedure -- significantly increases exposure over a dedicated self-initiated learning app. 

To critically reflect on this finding, we might ask if this embedding and increased exposure is ``worth it'' for users, considering potential downsides such as distraction or longer unlock interactions. Based on our results and user feedback here, we conclude that the overall answer is yes, yet our insights point towards individual differences: Both embedding approaches (notification, unlock) were favoured by a considerable proportion of people, who provided various concrete reasons for their choices. 

In the broader context of the literature, the results from our direct comparison here thus support such embedding and micro-learning concepts, and therefore strengthen their motivation and this line of research overall. Practically speaking, our findings motivate that language learning apps should consider offering these embedding concepts to users, for example, as complementary widgets to their main app content. 

Looking ahead, our findings encourage further investigations in this direction: For example, future work could explore the design space of unlock integration in detail (e.g. supporting further learning tasks and unlock mechanisms), also to improve the individually perceived trade-offs of learning and disturbance (e.g. via personalised and context-aware adaptations). Also, learning tasks could be connected to other smartphone actions, such as giving additional purpose to less productive actions defined by the user (e.g. starting a social media app). %